\documentclass[12pt]{article}
\usepackage{a4}
\usepackage{a4}\usepackage{epsf}\usepackage{latexsym}
\usepackage{Abbreviations}
\begin{document}
\title{Spontaneous Symmetry Breaking and Reflectionless Scattering Data}
\author{
  {\sc M. Bordag}\thanks{e-mail: Michael.Bordag@itp.uni-leipzig.de} \\
  \small  University of Leipzig, Institute for Theoretical Physics\\
  \small  Augustusplatz 10/11, 04109 Leipzig, Germany\\[9pt]
  \small and\\[12pt]
  {\sc A. Yurov}\thanks{e-mail: artyom.yurov@mail.ru}\\
  \small   Kaliningrad State University, Theoretical Physics Department\\
  \small 14, Al.Nevsky St., Kaliningrad 236041, Russia}
 
\maketitle \begin{abstract} 
We consider the question which potentials in the action of a (1+1) dimensional
scalar field theory allowing for spontaneous symmetry breaking have quantum
fluctuations  corresponding to reflectionless scattering data. The general
problem of restoration from known scattering data is formulated and a number of
explicit examples is given. Only certain sets of reflectionless scattering
data correspond to symmetry breaking and all restored potentials are similar
either to the Phi**4-model or to the sine-Gordon model. 
\end{abstract}
\section{Introduction}
Quantum corrections to classical solutions like kinks
\cite{Dashen1974e,raja82b} and spontaneous symmetry breaking are a fields of
intensive study and have applications in many branches of theoretical physics
ranging from the Standard model to solid state. Recent interest appeared from
some subtleties connected with supersymmetry \cite{recent}. A number of models
is usually considered in this connection. The most popular ones are the
$\Phi^4$-model and the sine-Gordon model. They result in a scattering problem
for the quantum fluctuations with reflectionless potentials. As a result
calculations of quantum corrections to the mass become very explicite. In the
present paper we investigate the question which models result in a
reflectionless scattering potential.  The surprising result is that all of
them are very similar to the above mentioned ones.

The setup of the problem is as follows. We consider a scalar field $\Phi(x,t)$
in (1+1) dimensions with action
\be\label{S}  S[\Phi]=\frac12 \int dxdt \left( \left(\pa_\mu
    \Phi\right)^2+U\left(\Phi\right)^2\right).
\ee
If the squared potential, $U^2\left(\Phi\right)^2$, has two (or more) minima
of equal depth \ssb occurs and topological nontrivial kink solutions
$\Phi_k(x)$ exist. In order to calculate the quantum fluctuations $\eta(x,t)$
in the background of the kink one has to solve the scattering problem for the
potential $V(x)$ which appears from the second derivative $\delta^2 S[\Phi_k]
/ \delta \Phi_k^2(x)$ of the action, see below Eq. \Ref{V2}. In simple models
like the mentioned above this potential $V(x)$ is reflectionless.

In the present paper we try to describe all potentials $U\left(\Phi\right)$ in
\Ref{S} which correspond to a reflectionless scattering potential $V(x)$ and
calculate the corresponding classical energy \Ec and the quantum energy $E_{0}$
which is the ground state energy of the field $\eta$ in the background of
$\Phi_k(x)$. 

In calculating these quantities it is usually assumed that the potential
$U\left(\Phi\right)$ is given. After that one solves the scattering problem
related to $V(x)$ and calculates the energies \Ec and $E_{0}$. In the paper
\cite{Bordag:1995jz} the inverse approach had been proposed. One starts from
the solution of the scattering problem given in terms of the so called
scattering data $\{r(k), \beta_i,\kappa_i\}$ known since \cite{Faddeev:1963yc}
to be in a one-to-one correspondence with the potential $V(x)$ (for a
representation of these questions see \cite{ChadanSabatier1989} and references
therein). Here $r(k)$ is the reflection coefficient, $\kappa_i$ are the bound
state energies and $\beta_i$ are numbers connected with the normalization of
the bound state wave functions.  As shown in \cite{Bordag:1995jz} the ground
state energy can be expressed in a simple way in terms of the scattering data
even including the necessary \uv renormalization, see Eq. \Ref{E0r} below.  In
order to find the classical energy one has to restore the potential $V(x)$
from the scattering data. This is the so called inverse scattering problem
which was solved in terms of certain integral equations (see, again,
\cite{ChadanSabatier1989}). In this way, solving the inverse scattering
problem the classical energy can be calculated from the scattering data. In
\cite{Bordag:1995jz} it was shown how this procedure works on the simplest
example of reflectionless ($r(k)=0$) scattering data containing only one bound
state.
 
In the present paper we use this inverse approach to describe all potentials
$U\left(\Phi\right)$ corresponding to reflectionless scattering data and
having topologically nontrivial solutions allowing in this way for spontaneous
symmetry breaking. It turns out that not all scattering data correspond to
such potentials $U\left(\Phi\right)$ but only certain classes. So we can
formulate the reconstruction problem: find the mapping between scattering data
and potentials $U\left(\Phi\right)$ allowing for spontaneous symmetry
breaking.

A special consideration deserve the so called rational scattering data. Here
the reflection coefficient $r(k)$ is a rational function of $k$ thus given by
a finite number of parameters. For a rational $r(k)$ the inverse scattering
problem is known to have an explicite, algebraic solution (in a similar way as
in the reflectionless case)  and the classical energy can be obtained then by
integration. In addition, the rational scattering data form a dense subset in
the set of all scattering data. In this way, the inverse approach may provide
an approximation scheme for the general case. 

The paper is organized as follows. In the next section we consider soliton
potentials providing completely explicite formulas. In the third section we
consider scattering data given by two bound states. In the fourth
section we show how this can be generalized to the general reflectionless
case. Conclusions are given in the last section. We use units with
$\hbar=c=1$.

\section{Formulation of the reconstruction problem}
We consider s scalar field $\Phi$ with action $S[\Phi]$, Eq. \Ref{S}, in (1+1)
dimensions. Static solutions $\Phi(x)$ are subject to the equation of motion
$\Phi''(x)=U(\Phi)U'(\Phi)$ where the prime denotes differentiation with
respect to the argument. We assume that $U^2(\Phi)$ has at least two minima of
equal depth  and we are free to denote two neighbored ones by $\pm\Phi_{\rm
  vac}$. These fields, $\Phi(x)=\pm\Phi_{\rm
  vac}$, are the vacuum solutions. In case $\Phi_{\rm  vac}\ne0$ there
exist topological nontrivial solutions $\Phi_k(x)$ called kink solutions which
interpolate between the vacuum solutions by means of
$\Phi_k(x\to\pm\infty)=\pm\Phi_{\rm  vac}$. These solutions obey the Bogomolny
equations
\be\label{Be} \Phi'_k(x)=U\left(\Phi_k(x)\right)
\ee
and have the classical energy
\be\label{Ecl0}E_{\rm class}=\frac12
\int_{-\infty}^\infty dx \left(\left(\Phi'_k(x)\right)^2+U^2\left(\Phi_k(x)\right)\right)
\ee
which by means of Eq. \Ref{Be} can be written in the form
\be\label{Ecl}E_{\rm class}=\int_{-\infty}^\infty dx \
U^2\left(\Phi_k(x)\right). 
\ee
In order to have a finite energy of the kink we must assume   that the
potential $U(\Phi)$ is zero in its minima.

The quantization of the scalar field in the background of the kink  solution
by means of the shift
\be\label{dec}\Phi(x,t)=\Phi_k(x)+\eta(x,t)
\ee
delivers in the Gaussian approximation the action
\be\label{Seta}S_{\rm fluct.}[\eta]=\frac12\int dx \ dt \
\eta(x,t)\left(\pa_t^2-\pa_x^2+\mu^2+V(x)\right)\eta(x,t)
\ee
for the fluctuations where the potential $V(x)$ results from
\be\label{V2} \frac12\frac{\delta^2 U^2(\Phi)}{\delta
  \Phi^2}_{\Big| _{\Phi=\Phi_k}}=(U'(\Phi))^2+U(\Phi)U''(\Phi)\equiv \mu^2+V(x).
\ee
Here $\mu$ is defined from demanding $V(x\to\infty)=0$ and has the meaning
of being the mass of the fluctuating field $\eta(x,t)$. 

The one loop quantum corrections to the energy are given by a functional determinant. For a static background they can be quivalently formulated   in
terms of the ground state energy   $E_0$ of $\eta(x,t)$ in the background of the kink,
\be\label{E0}E_0=\frac12\sum_{(n)}\epsilon_{(n)},
\ee
where the $\epsilon_{(n)}$ are the one particle energies of the fluctuations.
They are eigenvalues of the corresponding Schr\"odinger equation
\be\label{Se}\left(-\pa_x^2+\mu^2+V(x)\right) \eta_{(n)}(x)= \epsilon_{(n)}^2
\eta_{(n)}(x).
\ee
Here, the index $(n)$ denotes the spectrum of the operator in the lhs
of Eq. \Ref{Se}. In fact, Eq. \Ref{E0} defines $E_0$ only
symbolically.  One has to subtract the Minkowski space contribution
and to perform the \uv renormalization. These procedures are by now
well known. We follow here the treatment in \cite{Bordag:1995jz}. For
a discussion of the relations to different renormalization schemes we
refer to \cite{BGNV} where, for instance, the equivalence of the
subtraction scheme based on the heat kernel expansion and the mass
renormalization with the 'no tadpole condition' had been shown.

In terms of the scattering data the renormalized ground state energy
can be written in the form \cite{Bordag:1995}
\bea\label{E0r}E_0&=&\frac{-1}{4\pi^2}\int_0^\infty \frac{dq \ q}{\sqrt{\mu^2+q^2}} 
\log\frac{q+\sqrt{\mu^2+q^2}}{\sqrt{\mu^2+q^2}-q}\log\frac{1}{1-r(q)^2} \\ && \nn
-\frac{1}{\pi}\sum_{i=1}^N\left(\kappa_i-\sqrt{\mu^2-\kappa_i^2} \ \arcsin\frac{\kappa_i}{\mu}\right).
\eea
Here, the $\kappa_i$ are the binding energies of the bound states in the
potential $V(x)$,
\be\label{Sce}\left(-\pa^2_x+V(x)\right)\eta_i(x)=-\kappa_i^2 \eta_i(x),
\ee
where the $\eta_i(x)$ are the corresponding eigenfunctions. These are bound
state wave functions and they are normalizable, $\int_{-\infty}^\infty dx \ 
\eta^2_i(x) <\infty$. The function $r(k)$ is the reflection coefficient and
both, $\kappa_i$ and $r(k)$ belong to the scattering data. It should be
underlined that in $E_0$, Eq. \Ref{E0r}, the \uv divergences are subtracted.
This resulted in this quite simple form because the heat kernel coefficients
could be expressed in terms of the scattering data\footnote{This is related to
  the fact that here the heat kernel coefficients are just the conservation
  laws of the Korteweg-de-Vries equation.}. A nice consequence which can be
read off from this formula is that the ground state energy is always negative.

As mentioned in the introduction, the problem of calculating quantum
corrections can be inverted. One starts from the scttering data and by
means of Eq. \Ref{E0r} the quantum corrections can be obtained by
simple integration. The price one has to pay is a more complicated
procedure to obtain the classical energy. One has to solve the inverse
scattering problem, i.e., one has to reconstruct the potential $V(x)$
from the scattering data. This problem had been intensively studied in
connection with the solution of nonlinear evolution equations in the
70ies. The last step in this procedure is then to restore the
potential $U(\Phi)$ from $V(x)$ using Eq. \Ref{V2} and finally to
calculate the classical energy from Eq. \Ref{Ecl}.

In following this general procedure we make use of Eq. \Ref{V2} and the
Bogomolny equation \Ref{Be}. Differentiating Eq. \Ref{Be} twice with respect
to $x$ we obtain
\be\label{Y}\Phi'''(x)=\left(\left(U'(x)\right)^2+ U(x)U''(x)\right)\Phi'(x).
\ee
By means of Eq. \Ref{V2} and with the notation $\eta(x):=\Phi'(x)$ we rewrite
this equation in the form
\be\label{Y1}\left(-\pa_x^2+V(x)\right)\eta(x)=-\mu^2 \eta(x).
\ee
This equation shows that the derivative of the kink is a bound state solution
of the scattering problem associated with the potential $V(x)$ and that the
mass $\mu$ of the fluctuating field $\eta(x,t)$ in Eq. \Ref{Seta} is the
corresponding binding energy, i.e., one of the $\kappa_i$'s in the scattering
data. Note that $\eta(x)$ in Eq. \Ref{Y1} cannot be a scattering solution
because in that case $\mu^2$ would be negative. The decrease of $\eta(x)$ for
$x\to\pm\infty$ is by means of
\be\label{ieta}\int_{-\infty}^\infty dx \ \eta(x)=
\int_{-\infty}^\infty dx \ \frac{d}{dx} \Phi_k(x)=
\Phi_k(\infty)-\Phi_k(-\infty)=2\Phi_{\rm vac}
\ee
connected with a finite vacuum solution.

In this way, if we know $\eta(x)$, the field $\Phi(x)$ is given by
\be\label{feta0}\Phi_k(x)=-\Phi_{\rm vac}+\int_{-\infty}^x d\xi \ \eta(\xi)
\ee
and we restored $\Phi_k(x)$ from $\eta(x)$. 
The potential $U(\Phi)$ can be restored simply as
\be\label{Ueta0}U(\Phi_k(x))=\eta(x).
\ee
Note that the potential $U(\Phi)$ can be restored only from the ground state
wave function of the scattering potential $V(x)$ because it is only this
function which does not have zeros. In case $\eta(x)$ vanishes for some finite
$x$, the function $U(\Phi_k(x))$ would do so in contradiction to our
assumption that two neighbored zeros correspond to $x\to\pm\infty$.

In this way, by means of equations \Ref{feta0} and \Ref{Ueta0} we obtained a
parametric representation of the potential $U(\Phi)$ in terms of the ground
state wave function $\eta(x)$. We note that this representation covers the
region with $\Phi\in[-\Phi_{\rm vac},\Phi_{\rm vac}]$. How to go beyond we
consider in the following sections.

There is a freedom in the parametric representation, Eqs. \Ref{feta0},
\Ref{Ueta0}. The ground state wave function, $\eta(x)$, which we obtain as a
solution of the inverse scattering problem is determined up to a
multiplicative factor, which has the meaning of the normalization of $\eta(x)$
only. So we are free to multiply the function $\eta(x)$ by a constant,
$\eta(x)\to \al \eta(x)$. After that we can assume $\eta(x)$ to be
normalized, $\int_{-\infty}^\infty dx \ \eta(x)=1$.
In doing so we express $\al$ from Eq. \Ref{ieta} as
\[\al=2\Phi_{\rm vac}.
\]
In this way the freedom in the normalization of $\eta(x)$ is expressed in
terms of $\Phi_{\rm vac}$. After this rescaling we rewrite Eqs. \Ref{feta0}
and \Ref{Ueta0} in the final form
\be\label{feta}\Phi_k(x)=-\Phi_{\rm vac}+
2\Phi_{\rm vac} \int_{-\infty}^x d\xi \ \eta(\xi)
\ee
and
\be\label{Ueta}U(\Phi_k(x))=2\Phi_{\rm vac} \ \eta(x).
\ee
Using the last line we obtain from Eq. \Ref{Ecl} the classical energy which is
the quantity we are interested in,
\be\label{Ecleta} E_{\rm class}=
4\Phi_{\rm vac}^2 \ \int_{-\infty}^\infty dx \  \eta^2(x).
\ee
By the pair of equations, Eq. \Ref{E0r} and \Ref{Ecleta}, we obtained the final
expressions relating the complete energy
\be\label{Eges} E= E_{\rm class} +E_0
\ee
to the scattering data. 

However, it should be noticed that this is merely a formal solution. We
restored $U(\Phi)$ for a restricted range of $\Phi$ only. We have to construct
a continuation to all values of $\Phi$ which must deliver a single valued
function $U(\Phi)$ having the necessary extrema in order to allow for
spontaneous symmetry breaking. The investigation of this property is the main
difficulty in the restoration problem.

We conclude this section with a discussion of the free parameters.  First of
all there are the scattering data which constitute a set of independent
parameters. Second, we have the vacuum solution, $\Phi_{\rm vac}$, which is in
fact the condensate of the field $\Phi$. As seen from the above formulas there
is no more freedom in the restoration process. Together with the uniqueness of
the restoration of $\eta(x)$ from the scattering data the above mentioned
parameters are the only independent ones. As for the dimensions we note that
$\Phi_{\rm vac}$ is dimensionless (we work in (1+1) dimensions) and that the
bound state levels $\kappa_i$ have the dimension of a mass. For reflectionless
scattering data these are the only dimensional parameters and a rescaling
$\kappa_i\to\la\kappa_i$ results in $E\to\la E$. In the remaining paper of the
paper we put the mass scale equal to one. 
\section{Reconstruction from Soliton Potentials}
In this section we consider the case of reflectionless scattering data
($r(k)=0$) given by $N$ bound states with energy levels
\be\label{sollab}\kappa_i=i \qquad (i=1,2,\dots,N).
\ee
Here the ground state is that with number $i=N$.   
The potential $V(x)$ belonging to these scattering date is well known,
\be\label{VN}V(x)=\frac{-N(N+1)}{\cosh^2x}.
\ee
The solutions $\eta(x)$ of Eq. \Ref{Sce} are well known too. The ground state
wave function reads
\be\label{etaN}\eta(x)=\frac{1/\gamma_N}{\cosh^Nx}
\ee
and the corresponding eigenvalue is $\kappa_N=N$. The normalization factor
$\gamma_N$ is defined from $\int_{-\infty}^\infty dx \ \eta(x)=1$ and will be
calculated later in Eq. \Ref{gan}.  We call these $V(x)$ soliton potentials
because they are related to the soliton solutions of the Korteweg-de-Vries
equation.

\begin{figure}
[h]\unitlength=1cm
\begin{picture}(5,7.5)
\put(1,0){\epsfxsize=11cm \epsffile{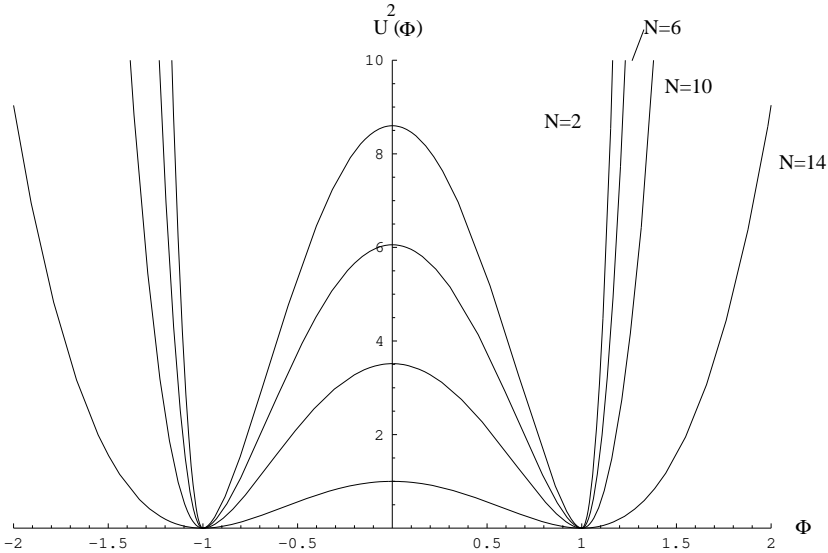}}
\end{picture}
\caption{The squared  potential $U^2(\Phi)$ reconstructed from a soliton
  potential with even number of bound states, $N=2,6,10,14$, and $\Phi_{\rm
  vac}=1$.
}
\label{figure1}
\end{figure}

Now, in order to solve the restoration problem we first consider even $N$. Here
it is useful to  change the  variable in Eq. \Ref{feta} according to 
\be\label{para1} x={\rm arctanh} \  t.
\ee
We introduce the notation $\Phi(t)=\Phi(x(t))$.  After that the integral over
$\xi$ in Eq. \Ref{feta} can be calculated easily and we arrive at
\bea\label{restf}\Phi(t)&=&-\Phi_{\rm  vac}+\frac{2\Phi_{\rm  vac}}{\gamma_N}
 \int_{-1}^t\frac{d\tau}{1-\tau^2} \ (1-\tau^2)^{N/2},\nn \\
&=&-\Phi_{\rm
  vac}+\frac{2\Phi_{\rm  vac}}{\gamma_N} \sum_{i=1}^{\frac{N}{2}-1}\left({\frac{N}{2}-1 \atop
  i}\right)\frac{(-1)^i}{2i+1}\left(t^{2i+1}+1\right),\\[5pt]
U(\Phi(t))&=& \frac{2\Phi_{\rm  vac}}{\gamma_N} \ (1-t^2)^{N/2}.\label{restU}
\eea
Now we observe that for $t\in[-1,1]$, or equivalently, for $x\in
(-\infty,\infty)$ we restored just the kink solution, $\Phi_k(t)$ and the
potential $U(\Phi_k(t))$ in a parametric representation. In this way we know
$U(\Phi)$ for $\Phi\in[-\Phi_{\rm vac},\Phi_{\rm vac}]$. However, the
parametrization \Ref{para1} together with the explicite formulas \Ref{restf}
and \Ref{restU} allow us to go beyond the region $t\in[-1,1]$. Simply, we
have to consider Eqs. \Ref{restf} and \Ref{restU} for $|t|>1$. For that $t$,
the variable $x$ becomes complex but $\Phi(t)$ and $U(\Phi(t))$ remain real.
We have to ensure that $t\in (-\infty,\infty)$ covers the whole range $\Phi\in
(-\infty,\infty)$ and that the resulting $U(\Phi)$ is a single valued
function. For this end we consider the derivative
\[ \frac{d\Phi(t)}{dt}=
\frac{2\Phi_{\rm  vac}}{\gamma_N} \ (1-t^2)^{\frac{N}{2}-1}.
\]
It may change its sign in $t=\pm 1$. If it changes it sign the function
$\Phi(t)$ is not monotonous and, as a consequence, $U(\Phi)$ is not single
valued.  If, in contrary, there is no change in the sign, $\Phi(t)$ is
monoton. Finally from the remark that $\Phi(t)$ is a polynomial in $t$ the
coverage of the whole region for $\Phi$ follows. This is the case for
$N=2(2s+1)$, $(s=1,2,\dots)$.  From Eq. \Ref{restU} it is seen that $U(\Phi)$
is in that case a function with two minima like in the $\Phi^4$-model.
For large $\Phi$, the asymptotic behavior is 
\[U(\Phi)\sifu \Phi^\frac{N}{N-1} \ .
\]
Some examples for $U(\Phi)$ are
shown in Fig.  \ref{figure1}.

For $N=2$ we reobtain the  $\Phi^4$-model. Here the explicite formulas
read 
\beao \Phi(t)&=& \Phi_{\rm vac} \ t,\\
U(\Phi(t))&=&\Phi_{\rm vac} \ ( 1-t^2),
\eeao
which can be trivially resolved,
\[
U(\Phi)=
\Phi_{\rm vac}\left( 1-\left(\frac{\Phi^2}{\Phi_{\rm vac}}\right)^2\right).
\]
The next example is $N=6$. Here the parametric representation reads
\beao \Phi(t)&=& \frac18 \Phi_{\rm vac} \ t\left(15-10 t^2+3 t^4\right),\\
U(\Phi(t))&=&  \frac{15}{8} \Phi_{\rm vac} \ \left(1-t^2\right)^3,
\eeao
which for $t\in(-\infty,\infty)$ defines the complete dependence $U(\Phi)$.
However, as can be seen, there is no explicite expression for $U(\Phi)$. Only
the inverse function can be given explicitely,
\[ \Phi(U)=\Phi(t)_{\Big|_{t=\sqrt{1-(8U/15\Phi_{\rm vac})^{1/3}}}},
\]
where the branches have to be chosen accordingly (the parametric
representation is much simpler). 

In this example we see explicitely how the continuation beyond the initial
region works. The reason that it works at all is that we assumed the
potential $U(\Phi)$ to be a function of $\Phi$ and not a more general object
like, for instance, a functional. 

\begin{figure}
[h]\unitlength=1cm
\begin{picture}(5,7.5)
\put(1,0){\epsfxsize=11cm \epsffile{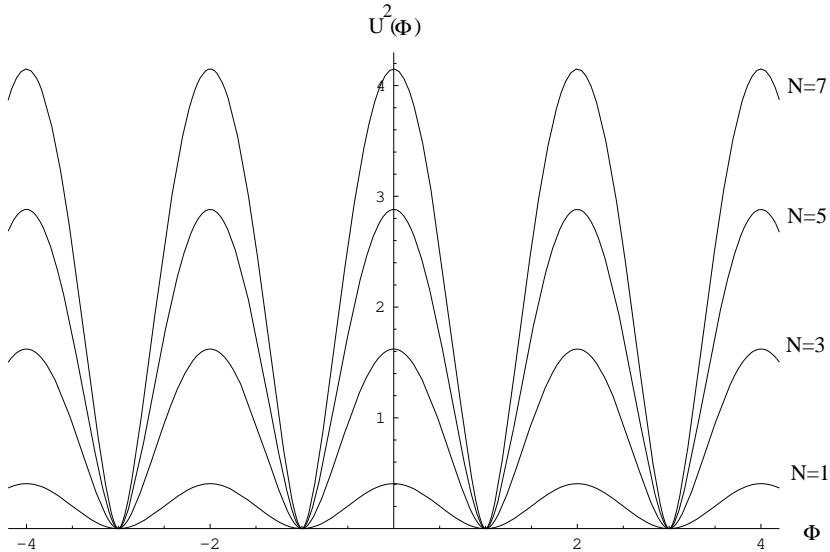}}
\end{picture}
\caption{The squared  potential $U^2(\Phi)$ reconstructed from a soliton
  potential with odd number of bound states, $N=1,3,5,7$, and $\Phi_{\rm
  vac}=1$. }
\label{figure2}
\end{figure}

Now we turn to odd $N$. Here it is useful to change the variable $x$ for
$\theta$ according to
\be\label{para2}\frac{1}{\cosh x}=\cos \theta.
\ee
We obtain again an explicite parametric representation,
\bea\label{restf2} \Phi(\theta)&=&\frac{\Phi_{\rm vac}}{\gamma_N}
 \left({N-1\atop \frac{N-1}{2}}\right) \ \frac{\theta}{2^{N-1}} \nn \\
&& +\frac{2\Phi_{\rm vac}}{\gamma_N} \sum_{k=0}^{(N-3)/2} \frac{1}{2^{2k-1}} \left({N-1\atop k}\right)
 \frac{\sin (N-1-2k)\theta}{N-1-2k}\nn \\
U(\Phi(\theta))&=&\frac{\Phi_{\rm vac}}{\gamma_N} \cos^N\theta.\label{restU2}
\eea
The region $x\in(-\infty,\infty)$ corresponds to
$\theta\in[-\frac\pi2,\frac\pi2]$ and \Ref{restf2} gives for that $\theta$ the
kink solution $\Phi_k(\theta)=\Phi_k(x(\theta))$. Again, we obtain from this explicit parametric
representation all $\Phi$ by going beyond this region to $|\theta|>\frac\pi2$.
From Eqs.  \Ref{restf2} and \Ref{restU2} it is obvious that $U(\Phi)$ defined
in this way is a single valued function. It has neighbored zeros located in
$\Phi=\pm\Phi_{\rm vac}$. It is a periodic function with
period $2\Phi_{\rm vac}$. So we see that for each odd $N$ the restoration
delivers a periodic potential $U(\Phi)$. For $N=1$ we note
\beao \Phi(\theta)&=&\frac{2\Phi_{\rm vac}}{\pi} \  \theta,\\
U(\Phi(\theta))&=&\frac{2\Phi_{\rm vac}}{\pi} \ \cos\theta,
\eeao
which can be resolved to $U(\Phi)=\frac{2\Phi_{\rm
    vac}}{\pi}\cos\left(\frac{\pi\Phi}{2\Phi_{\rm vac}}\right)$ which is the
sine-Gordon-model. For $N=3$ we obtain
\beao \Phi(\theta)&=& 
\frac{\Phi_{\rm vac}}{\pi} \left(2 \theta +  \sin(2\theta)\right),\\
U(\Phi(\theta))&=& \frac{4\Phi_{\rm vac}}{\pi}  \cos^3\theta.
\eeao
Again, there is an explicite expression for $\Phi(U)$ but no for $U(\Phi)$. 
Examples for some first odd N are given in Fig. \ref{figure2}.

\begin{figure}[h]\unitlength=1cm
\begin{picture}(5,7.5)
\put(1,0){\epsfxsize=11cm \epsffile{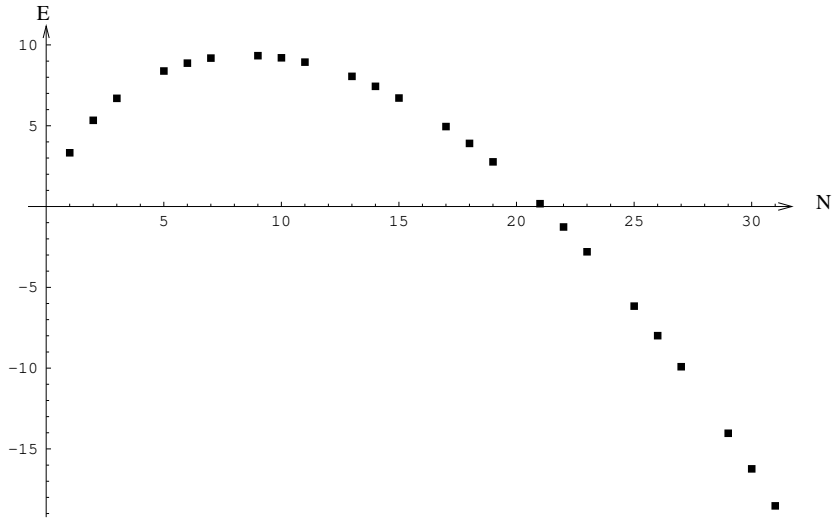}}
\end{picture}
\caption{The complete energy for soliton potentials with $N$ bound states,
the value of the condensate is  $\Phi_{\rm vac}=1.5$. }
\label{figure3}
\end{figure}

It remains to calculate the corresponding energies. The normalization factor
$\gamma_N$ in Eq. \Ref{etaN}  can be calculated explicitely,
\be\label{gan} \gamma_N=
\int_{-\infty}^\infty dx \ \frac{1}{\cosh^N x}=
\frac{\sqrt{\pi} \ \Gamma\left(\frac{N}{2}\right)}{\Gamma\left(\frac{(N+1)}{2}\right)}.
\ee
The asymptotics for large $N$ is $\gamma_n\sim \sqrt{\pi/(2N)}$.
Further we note
\[\int_{-\infty}^\infty dx \ \eta^2(x)=\gamma_{2N}.
\]
In this way we obtain
\be\label{Ecl1} E_{\rm class}=4\Phi_{\rm
  vac}^2\frac{\gamma_{2N}}{\left(\gamma_N\right)^2}
\ee
and 
\be\label{E01} E_0=
-\frac{1}{\pi}\sum_{i=1}^N\left(i-\sqrt{N^2-i^2} \ \arcsin\frac{i}{N}\right). 
\ee
As mentioned in \cite{Bordag:1995}, the renormalized vacuum energy is always
negative in (1+1) dimensions which can be checked for Eq. \Ref{E01} easily.
The classical energy is of course positive so that these two contributions to
the complete energy compete. For any finite $N$ it depends on $\Phi_{\rm
  vac}$ which prevails. For large $\Phi_{\rm vac}$ which correspond to a weak
coupling we have positive complete energy whereas for large $N$ the quantum
energy grows faster than the classical one. This is shown in Fig.
\ref{figure3}.
\section{Reconstruction from two bound states}
In this section we consider reflectionless scattering data consisting of two
bound states,
\beao \kappa_1&=&N_1,\\\kappa_2&=&N_2, \qquad \mbox{(ground state)}
\eeao
assuming $N_2>N_1$. The ground state wave function reads
\be\label{eta2l}\eta(x)=
\frac{2\cosh(N_1x)}{(N_2-N_1)\cosh((N_2+N_1)x)+(N_2+N_1)\cosh((N_2-N_1)x)} 
\ee
(up to the normalization factor). By means of Eqs. \Ref{feta} and \Ref{Ueta}
we restore $U(\Phi_k(x))$ and $\Phi_k(x)$. In this way we obtain information
on $U(\Phi)$ for $\Phi\in[-\Phi_{\rm vac},\Phi_{\rm vac}]$. To go beyond this
region we used in the preceeding section some specific parametrization. In
fact we made an analytic continuation to complex $x$. Indeed, for $|t|>1$ we
note for the first parametrization, Eq. \Ref{para1},
\be\label{fort1}x=\frac12\ln\frac{1+1/t}{1-1/t}\pm i\frac\pi2
\ee
and for the second one, Eq. \Ref{para2},  for
$\theta\in[\frac\pi2,\frac{3\pi}{2}]$ (where $\cos\theta<0$)
\be\label{fort2}x=
\ln\left(\frac{-1}{\cos\theta}-\sqrt{\frac{1}{\cos^2\theta}-1}\right)\pm i \pi.
\ee
Here the signs of the imaginary parts depend on which side we bypass the
corresponding branch point.  Aimed by these examples we consider $\eta(x+iy)$
(with real $x$ and $y$). Now we have to ensure that both, $U$ and $\Phi$ are
real.  Because $\Phi$ contains an additional integration as compared to $U$ we
need $\eta(x+iy)$ to be real for all $x$. Hence, only shifts in parallel to
the real axis are allowed. From the structure of $\eta$, Eq. \Ref{eta2l}, it
is clear that this may happen only if $N_1$ and $N_2$ are integer numbers and
if we take the shift in multiples of $i\frac\pi2$. In general, rational number
are possible too. But the denominators can be removed by a rescaling of $x$,
i.e. they can be absorbed into the mass scale.  In this way we see that the
two parametrization introduced in the preceeding section provide just the
required continuation.

As already mentioned we have to ensure that the parametrizations provide
monotone functions $\Phi(t)$ resp. $\Phi(\theta)$ which cover the whole range
$\Phi\in(-\infty,\infty)$. First we check the monotony. For that task we
consider the derivative of $\Phi$ with respect to the parameter. In the first
parametrization we note $dx/dt=1/(1-t^2)$ and obtain
\be\label{fit}\frac{d\Phi(t)}{dt}=\frac{\eta(x(t))}{1-t^2}
\ee
which must have a definite sign. A change in the sign may occur only in
passing through $t=1$, i.e., when going through $x\to\infty$. Using
\be\label{asU}
U(x)\sif \ e^{-N_2x}
\ee
and
\[x(t)\sift -\frac12\ln(1-t)
\]
we obtain
\be\label{asft}\frac{d\Phi(t)}{dt} \sift(1-t)^{\frac{N_2}{2}-1}.
\ee
This derivative is nonnegative for $t>1$ too only if $N_2=2(2s+1)$
($s=0,1,2,\dots$). 

In the second parametrization we have to investigate the behavior in
$\theta=\frac\pi2$. By means of $dx/d\theta=1/\cos\theta$ and
\[x(\theta)\sife -\ln\left(\frac\pi2 -\theta\right)
\]
we obtain
\be\label{asth}\frac{d\Phi(\theta)}{d\theta}=\frac{U(x(\theta))}{\cos\theta}
\sift \left(\frac\pi2 -\theta\right)^{N_2-1}
\ee
which is positive for $\theta>\frac\pi2$ for odd $N_2$, $N_2=2s+1$
($s=0,1,2,\dots$). 

In this way we arrived with the result that for each second even $N_2$ by the
first parametrization and for each odd $N_2$ by the second parametrization a
monotone function $\Phi(t)$ resp. $\Phi(\theta)$ appears. It remains to check
that the whole region $\Phi\in(-\infty,\infty)$ is covered. For the second
parametrization this is indeed the case simply by periodicity. However for the
first one this turns out not to be the case for all even $N_2$. To check this
we note that for $t\to\infty$ the real part of $x$ returns to zero as follows
from Eq. \Ref{fort1}. In $\eta(x)$, Eq. \Ref{eta2l}, after $x\to x+iy$, the
$\cosh$'s in the denominator turn into $\pm\sinh$'s of the corresponding
arguments. As a consequence, for $x\to0$ there may be a cancellation of the
contributions linear in $x$. It is just this cancellation which lets $U(x)$
grow up. It can be checked that this cancellation happens just for
$N_2=2(2s+1)$, i.e., for that we selected from the sign of the derivative, and
not for the other even $N_2$. There is no restriction on $N_1$. As a result we
obtain that the potential $U(\Phi)$ is again similar to that in the
$\Phi^4$-model, its asymptotic behavior is $U(\Phi)\sifu \Phi^2$.
 
\begin{figure}[h]\unitlength=1cm
\begin{picture}(5,14)
\put(1,8){\epsfxsize=11cm \epsffile{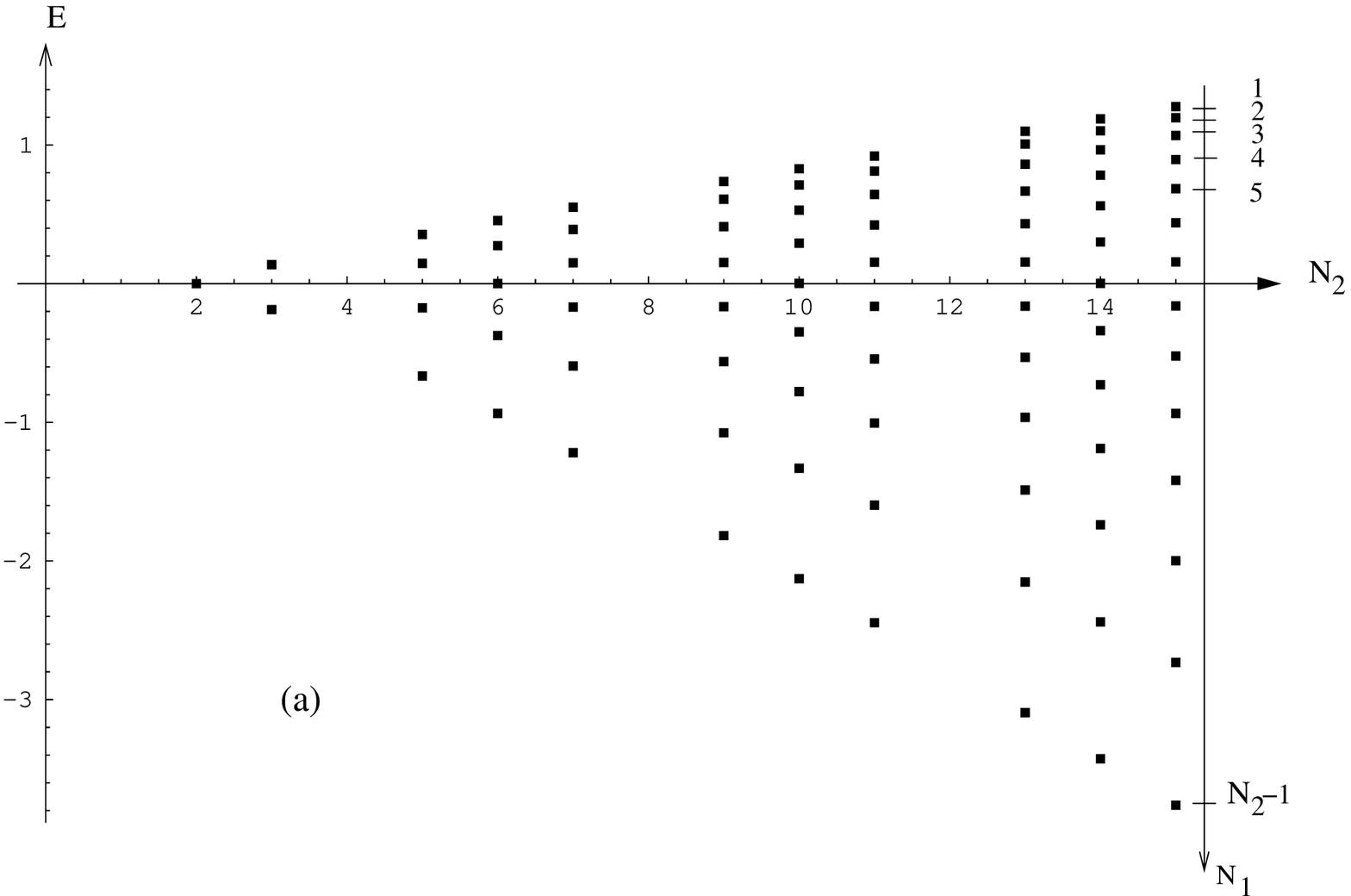}}
\put(1,0){\epsfxsize=11cm \epsffile{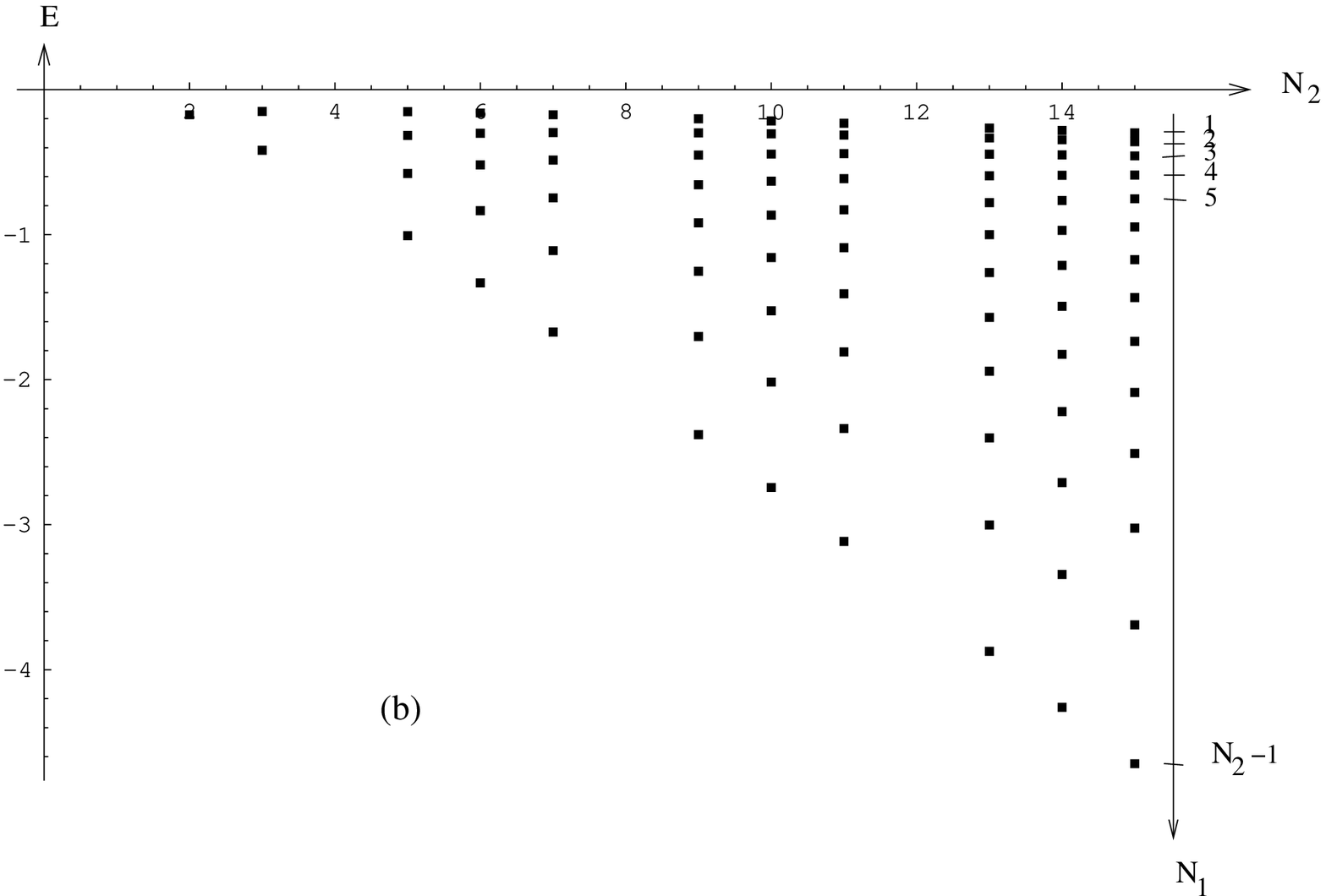}}
\end{picture}
\caption{The complete energy for potentials restored from two bound states,
  the value of the condensate is (a), $\Phi_{\rm vac}=0.5$ and (b), $\Phi_{\rm vac}=0.45$. }
\label{figure4}
\end{figure}

The classical energy can be calculated using Eq. \Ref{Ecleta}. However there
is no such simple explicite formula as in section 3. Results are shown in
Fig. \ref{figure4}. As seen it depends on the value of the condensate which
contribution prevails. For $N_1$ close to $N_2$, for any fixed value of the
condensate, the energy becomes negative for sufficiently large $N_2$. 
\section{Reconstruction from a general reflectionless potential}
In this section we consider a general reflectionless potential. Is is given by
$M$ bound states with energies $\kappa_i=N_i$ ($i=1,2,\dots,M$). We assume
$N_1<N_2<\dots<N_M$. The wave function of the ground state (its energy is
$N_M$) can be obtained from the inverse scattering method or by
Darboux-transformation. It is a quotient
\[ \eta(x)=\frac{P}{Q},
\]
where $P$ is a monomial in
$\cosh\left(\left(N_1\pm N_2\pm\dots\pm N_{M-1}\right)x\right)$ and $Q$ is a monomial in
$\cosh\left(\left(N_1\pm N_2\pm\dots\pm N_{M}\right)x\right)$. $Q$ contains the
ground state energy $\kappa_M=N_M$ and $P$ doesn't. Following the discussion in the
preceeding section we conclude that all $N_i$ must be integer. For the
behaviour at $x\to\infty$ from the largest in module eigenvalue    
\[ \eta(x)\sif e^{-N_Mx}
\]
follows. Again, we conclude that for $N_M=2(2s+1)$ ($s=0,1,2,\dots$) using the
first parametrization, Eq. \Ref{para1}, we obtain a monotone function
$\Phi(t)$ and that for odd $N_M$ the second parametrization does the job.
Whereas the second parametrization covers the whole region of $\Phi$ by
periodicity, the first does this only for certain sets of numbers
$N_1,N_2,\dots,N_{M-1}$. Here it seems too hard or even impossible to give a
general rule other than in special cases. So, for example, for three bound
states ($M=3$) and a ground state energy $N_3=2(2s+1)$, the energy of the
second level, $N_2$, must be an odd number and that of the first level, $N_1$,
an even number. This is a conjecture from considering $N_3$ explicitely up to
20. For four bound states ($M=4$) some allowed combinations are shown in Table
1.

The general behavior of $U(\Phi)$ is the same as seen before. For the ground
state energy being an even number a potential like in the $\Phi^4$-model
appears and for an odd number it is periodic. It seems that for 
reflectionless scattering data  no other behavior of $U(\Phi)$ is possible. 

\begin{table}
\begin{tabular}{c|cccccc}
&$N_1$ & 1&2&3 &4 \\\hline
$N_2$&  \\  
2&&1\\
3&&0&0\\
4&&1&0&1\\
5&&0&0&0&0\\
\multicolumn{3}{c}{(even $N_3$)}\end{tabular} \hspace{1cm}
\begin{tabular}{c|cccccc}
&$N_1$ & 1&2&3 &4 \\\hline 
$N_2$&  \\  
2&&1\\
3&&1&0\\
4&&1&1&1\\
5&&1&0&1&0\\
\multicolumn{3}{c}{(odd $N_3$)}\end{tabular}
\caption{Allowed (1) and forbidden (0) combinations of the bound state levels
  for four bound states. This is independent on the ground state level, $N_4$.}
\end{table}
\section{Conclusions}
We formulated the reconstruction problem on how to get the potential $U(\Phi)$
allowing for spontaneous symmetry breaking in the action, Eq. \Ref{S}, for a
scalar field in (1+1) dimensions from the scattering data related to the
quantum fluctuation in the background of the corresponding kink solution. We
considered reflectionless scattering data and solved the reconstruction
problem explictly for some classes, for soliton potentials and for two bound
states. We gave a conjecture for the general reflectionless case.  It states
that $U(\Phi)$ reconstructed from reflectionless scattering data can be only
like a $\Phi^4$-potential, i.e., with two minima, or periodic like in the
sine-Gordon model.

It would be interesting to give a proof of this conjecture. Furthermore, it
would be interesting to consider scattering data including reflections, for
example with a rational reflection coefficient and to see how other than the
two mentioned types appear.

We wrote down the formulas for the classical and the quantum energies in terms
of the scattering data resp. the ground state wave function. In the considered
examples it is seen that in dependence on the free parameters, the complete
energy may take both signs. In general, by an increase of the bound state
energies the quantum energy (it is negative) grows faster than the classical
one and the complete energy becomes increasingly negative.

\section*{Acknowledgment}One of us (AY) thanks the Gottlieb Daimler  and Karl
Benz  foundation for financial support and Leipzig University for kind
hospitality.\\
The authors are greatly indebted to D. Vassilevich for essential and
helpful discussions.

\end{document}